\documentclass[aps,prl,twocolumn,amssymb,superscriptaddress,showpacs]{revtex4-1}
\usepackage{float}
\usepackage{xcolor}
\usepackage{color}
\usepackage{graphicx}
\usepackage{dcolumn}
\usepackage{bm}
\usepackage{color}
\usepackage{amsmath}
\usepackage{amsfonts}
\usepackage{natbib}
\usepackage{graphicx}
\usepackage{dcolumn}
\usepackage{bm}

\begin{document}

\title{Floquet Engineering of Magnetism in Topological Insulator Thin Films}

\author{Xiaoyu Liu}
\affiliation{State Key Laboratory of Low-Dimensional Quantum Physics, Department of Physics, Tsinghua University, Beijing 100084, China}
\affiliation{Max Planck Institute for the Structure and Dynamics of Matter and Center for Free-Electron Laser Science, Luruper Chaussee 149, 22761 Hamburg, Germany.}

\author{Peizhe Tang}
\email{peizhe.tang@mpsd.mpg.de}
\affiliation{Max Planck Institute for the Structure and Dynamics of Matter and Center for Free-Electron Laser Science, Luruper Chaussee 149, 22761 Hamburg, Germany.}

\author{Hannes H\"ubener}
\affiliation{Max Planck Institute for the Structure and Dynamics of Matter and Center for Free-Electron Laser Science, Luruper Chaussee 149, 22761 Hamburg, Germany.}

\author{Umberto De Giovannini}
\affiliation{Max Planck Institute for the Structure and Dynamics of Matter and Center for Free-Electron Laser Science, Luruper Chaussee 149, 22761 Hamburg, Germany.}

\author{Wenhui Duan}
\affiliation{State Key Laboratory of Low-Dimensional Quantum Physics, Department of Physics, Tsinghua University, Beijing 100084, China}
\affiliation{Institute for Advanced Study, Tsinghua University, Beijing 100084, China}
\affiliation{Collaborative Innovation Center of Quantum Matter, Beijing 100084, China}

\author{Angel Rubio}
\email{angel.rubio@mpsd.mpg.de}
\affiliation{Max Planck Institute for the Structure and Dynamics of Matter and Center for Free-Electron Laser Science, Luruper Chaussee 149, 22761 Hamburg, Germany.}
\affiliation{Nano-Bio Spectroscopy Group,  Departamento de Fisica de Materiales, Universidad del Pa\'{i}s Vasco UPV/EHU- 20018 San Sebasti\'{a}n, Spain.}
\affiliation{Center for Computational Quantum Physics, Flatiron Institute, 162 Fifth Avenue, New York, NY 10010, USA.}

\begin{abstract}
Dynamic manipulation of magnetism in topological materials is demonstrated here via a Floquet engineering approach using circularly polarized light. Increasing the strength of the laser field, besides the expected topological phase transition, the magnetically doped topological insulator thin film also undergoes a magnetic phase transition from ferromagnetism to paramagnetism, whose critical behavior strongly depends on the quantum quenching. In sharp contrast to the equilibrium case, the non-equilibrium Curie temperatures vary for different time scale and experimental setup, not all relying on change of topology. Our discoveries deepen the understanding of the relationship between topology and magnetism in the non-equilibrium regime and extend optoelectronic device applications to topological materials.
\end{abstract}

%would be interesting and important both for fundamental researches and practical applications. Herein, we report topological and magnetic properties in magnetic topological insulator (TI) thin films well tuned via the Floquet engineering.

%Furthermore, we propose an all-optical transistor device in which on-off signal from polar Kerr-rotation angel can be well controlled by the laser field.
\maketitle

%\tableofcontents

%\section{\label{sec:level1}Introduction}
The interplay between magnetism and topological state attracts the considerable attention in recent studies. Many exotic topological particles and macroscopic quantum phenomena have been realized, including Weyl fermions \cite{Armitage2018,Gang2011,yang2017topological,liu2018giant,kuroda2017evidence,ChangGQ2017}, antiferromagnetic Dirac fermions \cite{tang2016dirac,JingWang2017}, quantum anomalous Hall (QAH) effect \cite{yu2010quantized,chang2013experimental,Kou2014,checkelsky2014trajectory,FengYang2015,chang2015high,Gang2015}, antiferromagnetic topological insulators \cite{Mong2010,li2018intrinsic}, axion insulators \cite{li2010dynamical,WanXG2012,mogi2017magnetic,DiXiao2018,zhang2018topological,li2018intrinsic}. In real materials, the emergence of magnetism breaks the time reversal symmetry (TRS) and the exchange coupling can drive the system into a new topological phase. For example, the long-range ferromagnetic (FM) order can be formed by doping topological insulator (TI) films with magnetic ions \cite{yu2010quantized,LiuQ2009,checkelsky2012dirac,ChangCZ2013,zhang2013topology}, the Zeeman coupling splits degenerate bands and drives TI thin film to be a Chern insulator with the quantized edge conductance \cite{chang2013experimental,Kou2014,checkelsky2014trajectory,FengYang2015,chang2015high}. Meanwhile, in turn, these topological bands, both in bulk and on surface, have the ability to form or enhance long-range ferromagnetism \cite{yu2010quantized,zhang2013topology,LiuQ2009,checkelsky2012dirac}. Therefore, from a practical point of view, finding an efficient way to manipulate magnetic and topological orders will trigger intense researches in the fields of spintronics and quantum information processing. The electric field controlled ferromagnetism has been proposed theoretically \cite{wang_electrically_2015} and achieved experimentally in a magnetically doped topological insulator (MDTI) thin film \cite{zhang2017magnetic}, which is possible to be used in spintronic devices \cite{wang_electrically_2015}.

Different from conventional manipulations of topology and magnetism in equilibrium, the Floquet engineering via ultrafast optical driving provides a new possibility to tune electronic properties of host material dynamically in non-equilibrium \cite{oka2018floquet,hubener2017creating,hannes2018phonon,shin2018phonon,shin2019unraveling}. The strong interaction of photons with charges, spins and angular momentums can drive the system to a steady Floquet state, dramatically changing Floquet band structure and charge population \cite{oka2018floquet,hannes2018phonon,shin2018phonon}. This approach provides a new degree of freedom to manipulate the magnetic order \cite{Andrei2010,buzzi2018probing} and the topological phase \cite{Kitagawa2010,wang2013observation,mahmood2016selective,hubener2017creating,WangZF2018,LiuHS2018,oka2018floquet}. By using Floquet engineering, semiconductor quantum wells \cite{lindner2011floquet,dalessio_dynamical_2015,Seetharam2015} and two dimensional (2D) materials \cite{WangZF2018,claassen2016all} can be driven to behave as Floquet TIs. Light-induced Hall conductance is predicted and observed in graphene, pointing towards a topological origin \cite{oka2009,dehghani_dissipative_2014,dehghani_out--equilibrium_2015,ZhaiXC2014,Perez2014,sentef2015theory,mciver2018light,Sato2019}.

Herein, in the framework of the Floquet theorem, we study the electronic and magnetic properties of periodically driven TI and MDTI thin films. Under the circularly polarized light (CPL), the Floquet bands of steady states can be well tuned, resulting in a photon-induced topological phase transition (PT). Furthermore, we develop a Floquet linear-response theory to calculate the magnetic susceptibility in MDTI thin films. We find that CPL can reduce its magnetic susceptibility, driving FM thin films to be paramagnetic (PM). The decreasing behavior of the magnetic PT strongly depends on quantum quenching and dissipation. For example, in contrast to that in long time scale, the magnetic susceptibility decreases more quickly in short time scale  with the change of the laser field, resulting in a lower Curie temperature ($T_C$) correspondingly. Our work provides a new way to manipulate topological and magnetic properties in MDTI thin films via an optical approach and our predictions can be measured by magneto-optic and transport experiments.

\begin{figure}
\includegraphics[width=1.0\columnwidth]{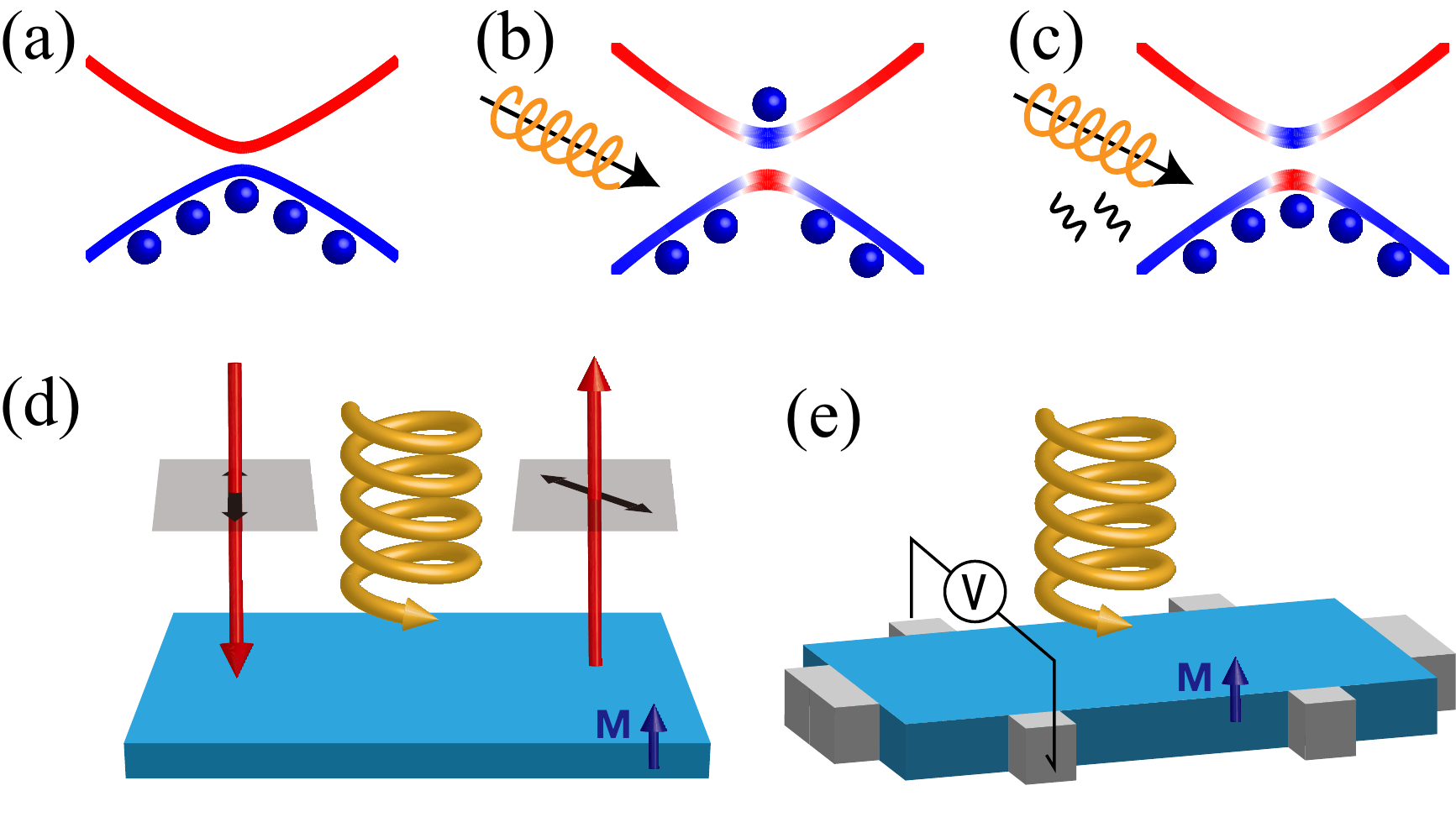}
	\caption{(a) Electronic bands and charge population in equilibrium. Floquet bands and charge population pumped by laser field with (b) the sudden approximation and (c) the Floquet Fermi-Dirac distribution. Red and blue lines represent states that originally belong to conduction and valance bands in equilibrium. Blue balls show the population of electrons schematically and black wavy lines stand for dissipation. Possible experimental setups are drawn schematically with (d) the magneto-optical Kerr rotation and (e) the ultrafast anomalous Hall transport. The magnetism is along our-of-plane direction. }\label{fig:diagram}
\end{figure}

\emph{Floquet theory}.-- Under CPL, a non-equilibrium \emph{Floquet state} is possible to be formed in quantum materials \cite{mahmood2016selective,wang2013observation,hubener2017creating,sentef2015theory}. In the Floquet theory \cite{Shirley1965,Sambe1973,grifoni1998driven}, we can solve the time-dependent Schr$\ddot{\rm o}$dinger equation in the Hilbert space $\mathcal{H}$ via mapping it to a $\emph{time-independent}$ infinite eigenvalue problem in an extended Hilbert space $\mathcal{H}\otimes \mathcal{L}_T$, \cite{Eckardt2015}\cite{SuppMater} where $\mathcal{L}_T$ represents the space of ``multi-photon-dressed" states. Finally, resulting states are the Floquet bands whose occupations strongly depend on the quantum quenching and dissipation details.

The band structure and charge distribution in initial equilibrium are shown in Fig. \ref{fig:diagram}(a). If we switch on the laser suddenly and measure the physical observable in a short time (such as magneto-optical Kerr rotation angle \cite{Kirilyuk2010}, schematic shown in Fig. \ref{fig:diagram}(d)), the time scale for this system to reach steady state is short as compared to the period of laser \cite{dehghani_dissipative_2014,dehghani_out--equilibrium_2015}. In such way (see Fig. \ref{fig:diagram}(b)), electrons tend to stay in their original bands, the population of the quenched Floquet system is mainly determined by the overlapping matrix between final and original states \cite{SuppMater}. This approach is named as the \emph{Sudden Approximation} \cite{dehghani_dissipative_2014,dehghani_out--equilibrium_2015}. On the other limit (see Fig. \ref{fig:diagram}(c)), if the measurement is performed in a long time (such as anomalous Hall effect in ultrafast transport \cite{mciver2018light,Sato2019}, schematic shown in Fig. \ref{fig:diagram}(e)), excited electrons are relaxed and the population of the Floquet state behaves like a \emph{Floquet Fermi-Dirac} distribution with effective chemical potentials for electrons and holes \cite{Iadecola2015,Seetharam2015}. In this work, we calculate magnetic properties, especially the magnetic susceptibilities for MDTI thin films in these two extreme situations by the Floquet linear response theory \cite{SuppMater,stefanucci_nonequilibrium_2013,dehghani_out--equilibrium_2015,oka2009,Chen2018}.

\emph{Effective Model}.-- To study MDTI thin films under CPL, we start from the 2D effective Hamiltonian for TI thin films \cite{zhang2010crossover,shan2010effective,wang_electrically_2015,yu2010quantized},
\begin{equation}\label{eq:thin_film_kp}
\begin{split}
  H_0(\textbf{k})= & \epsilon(\textbf{k})+v_F k_y \sigma_1\otimes\tau_3 -v_F k_x\sigma_2\otimes\tau_3\\
  & +m(\textbf{k})\cdot I\otimes\tau_1
\end{split}
\end{equation}
written in the basis of \{$|u\uparrow\rangle,|u\downarrow\rangle,|d\uparrow\rangle,|d\downarrow\rangle$\}, $\uparrow$ and $\downarrow$ stand for the spin degree of freedom, $|u\rangle$ and $|d\rangle$ are surface states on upper and down surfaces. $\sigma_i$ and $\tau_i$ (i=1,2,3) are the Pauli matrices for spin and orbital, $k_{x,y}$ is the momentum in reciprocal space, $v_F$ is the Fermi velocity, $\epsilon(\textbf{k})=\epsilon_0+\epsilon_2 (k_x^2+k_y^2)$ is the energy shift, and $m(\textbf{k})=m_0+m_2(k_x^2+k_y^2)$ is the coupling between $|u\rangle$ and $|d\rangle$. Due to the inter-surface coupling, a finite gap opens around the $\Gamma$ point and topological properties are determined by the sign of $m_0 m_2$ \cite{liu_model_2010,yu2010quantized,SuppMater}.

We include the effect of CPL by the Peierls substitution $\mathrm{\mathbf{k}}\rightarrow\textbf{k}-e\textbf{A}(t)$, where $\textbf{A}(t)$ is the laser amplitude, $e$ is the electron charge, and $\hbar$ is the reduced Planck constant. $\textbf{A}(t)=A(\cos\omega t,\sin\omega t,0)$ is for right-hand CPL, $\omega$ is the frequency with the photon energy of 3 eV. Time-independent infinite Floquet Hamiltonian $H_F(\textbf{k})$ can be obtained in the framework of the Floquet theory \cite{SuppMater}. By diagonalizing $H_F(\textbf{k})$ with a cut-off on the frequency domain, we obtain the Floquet band structures \cite{SuppMater}. In the high frequency limit ($\hbar\omega \gg W$, $W$ is the band width of thin film), the low order expansion \cite{Eckardt2015,oka2018floquet,mikami_brillouin-wigner_2016,hubener2017creating} can be employed to reduce the infinite $H_F(\textbf{k})$ to an effective Hamiltonian $H_{eff}(\textbf{k})$ with the form of
\begin{equation}\label{eq:2d_downfold}
\begin{split}
& H_{eff}(\textbf{k}) = H_0(\textbf{k}) +\epsilon_2 a^2 + m_2 a^2 \cdot I\otimes\tau_1\\
& +\frac{a^2}{\hbar\omega}v_F^2 \cdot \sigma_3\otimes\tau_0-2\frac{a^2}{\hbar\omega}v_F m_2(k_x\sigma_1+k_y\sigma_2)\otimes\tau_2
\end{split}
\end{equation}
Here $a=eA$ is the effective laser amplitude with unit \AA$^{-1}$. This Hamiltonian has the same dimension as the original $H_0(\textbf{k})$. Its second term corrects $\epsilon_0$ (see Eq. \ref{eq:thin_film_kp}) and results in the total energy shift. The third term modifies $m_0\rightarrow m_0+m_2 a^2$ and induces a topological PT when the laser amplitude is large enough. Interestingly, these terms do not depend on the laser frequency but rely on laser amplitude only. As shown in the second row of Eq. \ref{eq:2d_downfold}, CPL introduces two new terms, which are proportional to $a^2/\hbar\omega$ and smaller compared to correction terms under high-frequency limit. The first one is momentum independent and acts like the exchange coupling due TRS breaking. The second one is momentum dependent and has negligible effect around $\Gamma$.
To sum up, the change of band structures for TI thin films under CPL is dominated by correction terms and slightly modified by the frequency dependent terms.

\begin{figure}
\includegraphics[width=1.0\columnwidth]{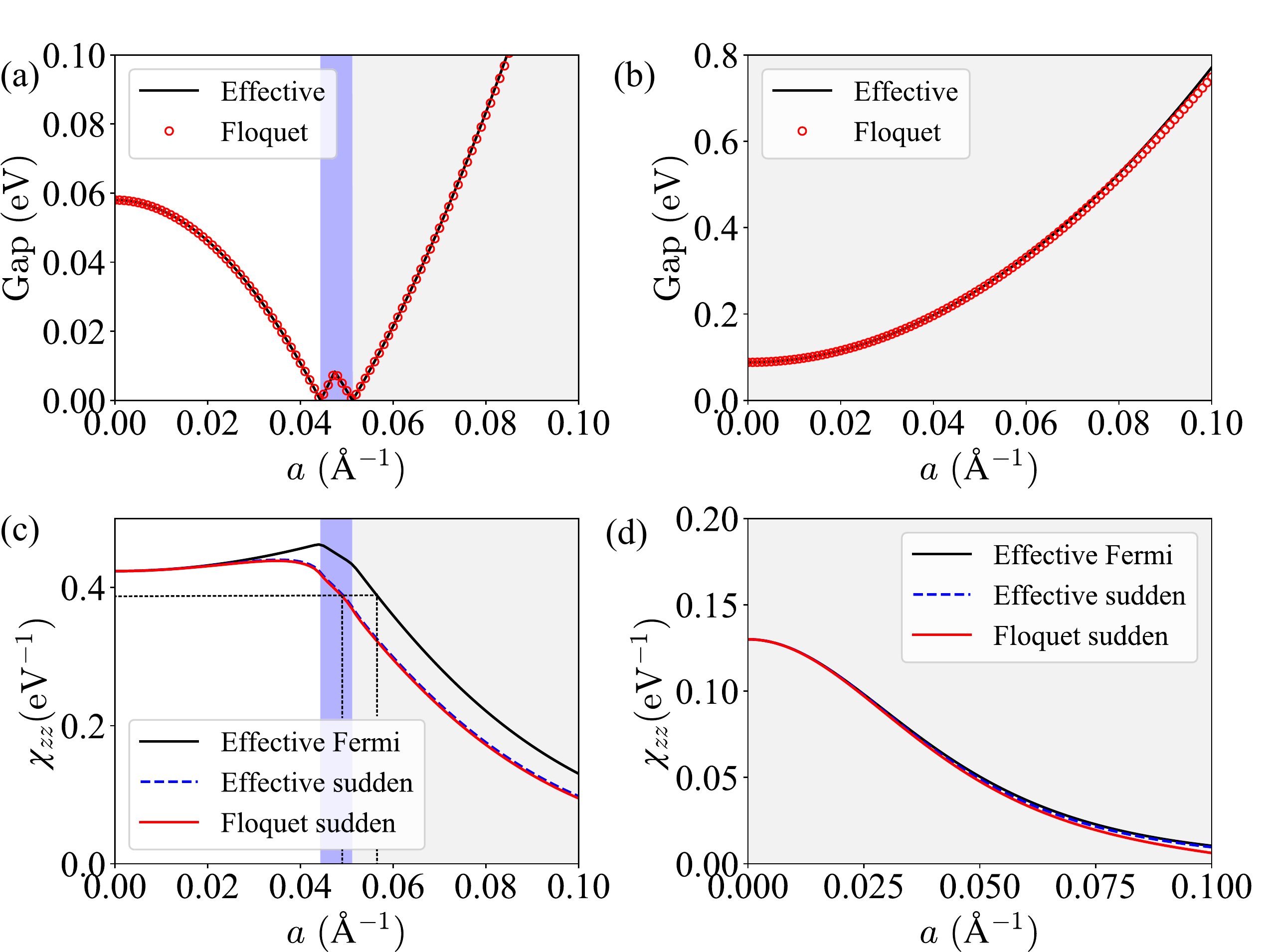}
\caption{ (a,b) Floquet gap size changes for TI thin films from the 2D effective model. In equilibrium, they are in topologically non-trivial (a) and trivial (b) phases. Black lines and red circles are obtained from the high frequency expansion and diagonalization of infinite Floquet Hamiltonian with a cut-off on the frequency domain. (c,d) The change of magnetic susceptibilities with $a$ in MDTI thin films. In equilibrium, they are in topologically non-trivial (c) and trivial (d) phases. Black solid lines and blue dashed lines are from the high frequency expansion with different population approximations. Red curves are results from diagonalization with sudden approximation. Background colors indicate topological phases and follow the convention shown in Fig. \ref{fig:2dphase}.}\label{fig:2d_result}
\end{figure}

Numerical results demonstrate the analysis above correct. In Fig. \ref{fig:2d_result}(a) and (b), we show the calculated direct Floquet gap size for TI thin films from the 2D effective model when fixing the frequency and changing the amplitude of the driving laser. In equilibrium, we use parameters shown in Ref. \cite{wang_electrically_2015} for $H_0(\textbf{k})$ where 4QL TI thin film is in the quantum spin Hall (QSH) phase ($m_0 m_2<0$) and 3QL thin film is a normal insulator ($m_0 m_2>0$). Turning on the laser, their electronic band structures are modified. Two curves with different color show the gap size in the Floquet band structures obtained from the high frequency expansion (black line) and direct diagonalization respectively (red circles). They agree very well, indicating that current parameter settings reach the high frequency limit.

When we increase the laser amplitude $a$ for 4QL thin film, its Floquet gap at the $\Gamma$ point decreases at first, goes through two gap-closing processes, and finally increases again (see Fig. \ref{fig:2d_result}(a)). Such evolution indicates topological PTs from a NI$B$ phase to a QAH phase, and finally to a NI$A$ phase. Here, NI$B$ is a state with Floquet band structure similar to the QSH insulator while not topologically protected \cite{wang_electrically_2015}. NI$A$ stands for a normal insulator. Herein, the QAH phase (blue region) mainly originates from the light-induced exchange coupling in TI film. Since the exchange term is proportional to $a^2/\hbar\omega$, its phase region is very narrow in the high frequency. In contrast to 4QL thin film, we observe no gap closing process as the increase of light amplitude for 3QL film, meaning that persists in NI$A$ phase (see Fig. \ref{fig:2d_result}(b)).

%Ferromagnetic order
When TI thin films are doped with magnetic ions, such as Cr \cite{chang2013experimental,ChangCZ2013} and V \cite{chang2015high} atoms, the long-range FM order can be formed in equilibrium. If we keep MDTI thin films insulating, the ferromagnetism is induced by the van Vleck mechanism \cite{yu2010quantized,zhang2013topology,ChangCZ2013,SuppMater}. The low energy electronic structure around the $\Gamma$ point can be described by the Hamiltonian: $H_0^{FM} (\textbf{k})=H_0(\textbf{k})+\Delta \cdot \sigma_3 \otimes I$. $\Delta$ is the Zeeman term arising from FM order and is determined by the density of dopants and the magnetic susceptibility \cite{yu2010quantized}. Interestingly, MDTI thin films undergo a FM to PM transition via changing dopant densities, accompanied by a topological PT \cite{zhang2013topology}.

We can fix the density of dopants in MDTI thin films to guarantee their equilibrium states to be FM for 4QL and PM for 3QL, and then apply CPL to drive the systems to the Floquet states. When timescales of magnetic impurity dynamics are much slower than $\omega^{-1}$, the time dependence of local moments on magnetic dopants can be neglected. Thus the change of magnetic properties induced by the laser field is determined by the magnetic susceptibilities of electrons $\chi_{zz}^{Floq}$, and it is calculated for Floquet states by the Kubo linear-response theory \cite{SuppMater}
\begin{equation}
{\chi_{zz}^{Floq}=\sum_{\alpha\in FBZ,\beta,k}2f_\alpha\frac{|\langle\langle\phi_\alpha(t)|s_z|\phi_\beta(t)\rangle\rangle|^2}{\epsilon_\beta-\epsilon_\alpha}\label{eq:chizz}}
\end{equation}
where $\alpha$ and $\beta$ label the Floquet states, $\phi_{\alpha}(t)$ is the Floquet wavefunction with quasienergy $\epsilon_{\alpha}$, $f_{\alpha}$ is its population, $s_z$ is the spin operator along $z$ direction, and FBZ is the Floquet Brillouin zone \cite{SuppMater}. In non-equilibrium, the occupation of Floquet states strongly depends on frequency and amplitude of driving laser \cite{ono_nonequilibrium_2018,Sato2019} and the dissipation to a bath \cite{Seetharam2015}. Here we only consider steady Floquet states formed in two limited situations with high frequency.

As shown in Fig. \ref{fig:2d_result}(c) and (d), when the light intensity goes up, the magnetic PT points change with different quenching schemes although the trend of their variation $\chi_{zz}^{Floq}$ behaves similarly. For 4QL MDTI thin film, the magnetic susceptibility increases slowly in the NI$B$ phase and drops quickly in the QAH and NI$A$ phases. When $\chi_{zz}^{Floq}$ reaches to the critical value resulting in a magnetic PT, we observe different critical points $a_c^{FM}$ for different quenching schemes (marked by black dashed line in Fig. \ref{fig:2d_result}(c)). Compared with $T_C$ in equilibrium, CPL decreases its value by the Floquet engineering. Such decreasing tendency is mainly contributed by the change of gap size and matrix elements shown in Eq. \ref{eq:chizz}.

In the regime of the sudden approximation, the critical value $a_c^{FM}$ is smaller than that in the Floquet Fermi-Dirac distribution. And both $a_c^{FM}$ in two kinds of quenched schemes do not have the same value as the critical amplitude $a_c^T$ for topological PT. Therefore, we conclude that, beside the modulation of energy levels, the light changes the population of Floquet states that also plays an essential role to determine the magnetic behavior. This population dependence is unique for non-equilibrium states in contrast to previous studies \cite{wang_electrically_2015,zhang2013topology}. For 3QL (see Fig. \ref{fig:2d_result}(d)), the decreasing behavior of $\chi_{zz}^{Floq}$ is observed but the whole system stays in PM phase.

%In NI$B$ phase, the reduction of gap size cause the denominator of Eq.(\ref{eq:chizz}) to be small, so the value The reduce of band inversion for 4QL MDTI films as light amplitude rises cause the numerator of Eq.(\ref{eq:chizz}) to reduce \cite{yu2010quantized}. Near topological PT, the latter dominates and we observe a sudden decrease of $\chi_{zz}$.

\begin{figure}
\includegraphics[width=0.95\columnwidth]{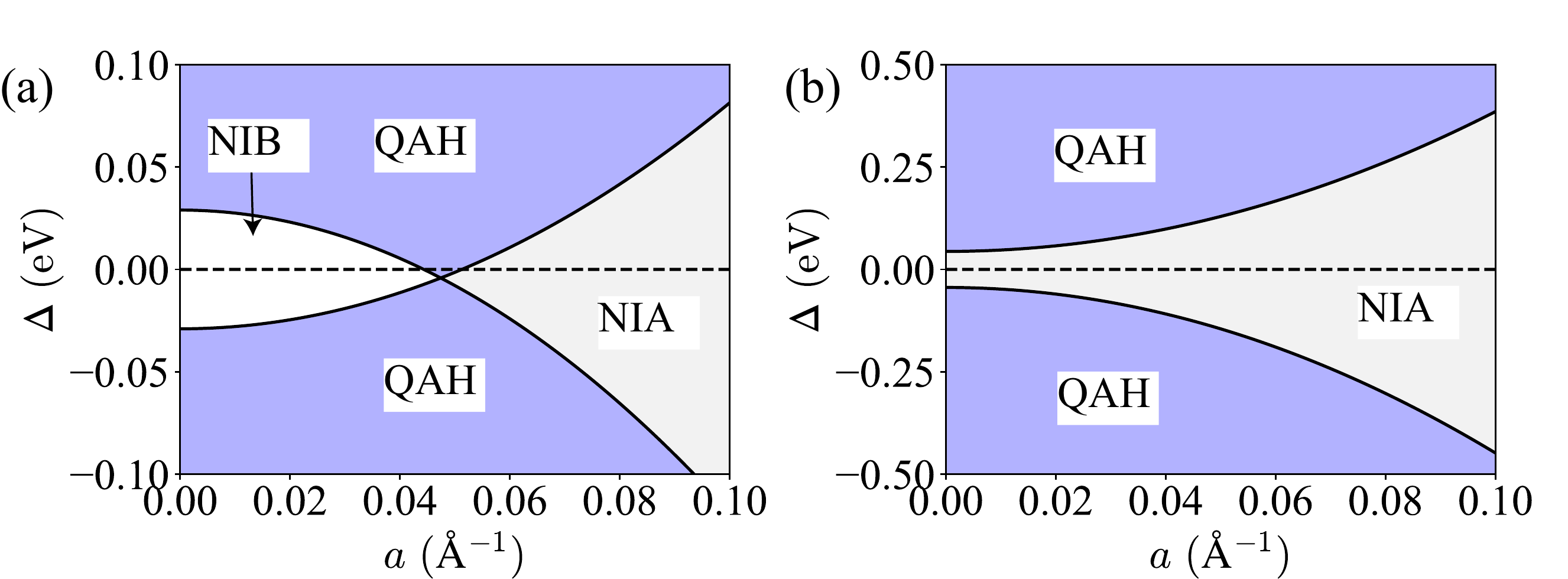}
\caption{Light induced topological phase diagram for TI ($\Delta=0$) and MDTI ($\Delta\neq 0$) thin films with the thickness of (a) 4QL and (b) 3QL. The QSH phase is well defined only at the point with $\Delta=0$ and $a=0$.}\label{fig:2dphase}
\end{figure}

The topological phase diagram is shown in Fig. \ref{fig:2dphase} for MDTI thin films, whose results are from the effective model considering the FM exchange coupling and the influence of CPL. If the intrinsic system is initially in QSH state (e.g. 4QL), there are three phase regions as shown in Fig. \ref{fig:2dphase}(a), including a QAH phase and two trivial insulating phases NI$A$ and NI$B$. The asymmetric feature with respect to $\Delta$ is due to the additional exchange terms induced by the driving laser (See Eq. \ref{eq:2d_downfold}). For an initial state without topology (e.g. 3QL), the phase diagram is much simpler with only transitions from the normal insulator to QAH state when $\Delta$ is large enough.

\emph{Thin film model}.-- The above discussions are based on the effective Hamiltonian. It is a relatively simple model to give us a clear physical picture of light induced magnetic and topological PTs, but cannot predict all details precisely without including the contributions from quantum well states (QWSs). To confirm the validity of the picture and describe a more realistic situation in experiments, we choose $(\mathrm{Sb}_{0.9}\mathrm{Bi}_{0.1})_2 \mathrm{Te}_3$ as an example that has been fabricated in previous experiments \cite{zhang2017magnetic,zhang2010crossover}. We consider its three-dimensional bulk Hamiltonian in a thin film construction with thickness $d$ \cite{SuppMater}. The confinement in the $z$ direction quantizes the momentum on this axis. The contribution from QWSs can be fully considered in thin film model \cite{wang_electrically_2015,zhang_ti_2009,liu_model_2010,liu_oscillatory_2010}. In our equilibrium calculations, $(\mathrm{Sb}_{0.9}\mathrm{Bi}_{0.1})_2 \mathrm{Te}_3$ thin film with 6QL is topologically trivial and its 5QL thin film is a QSH insulator.

\begin{figure}
\includegraphics[width=1\columnwidth]{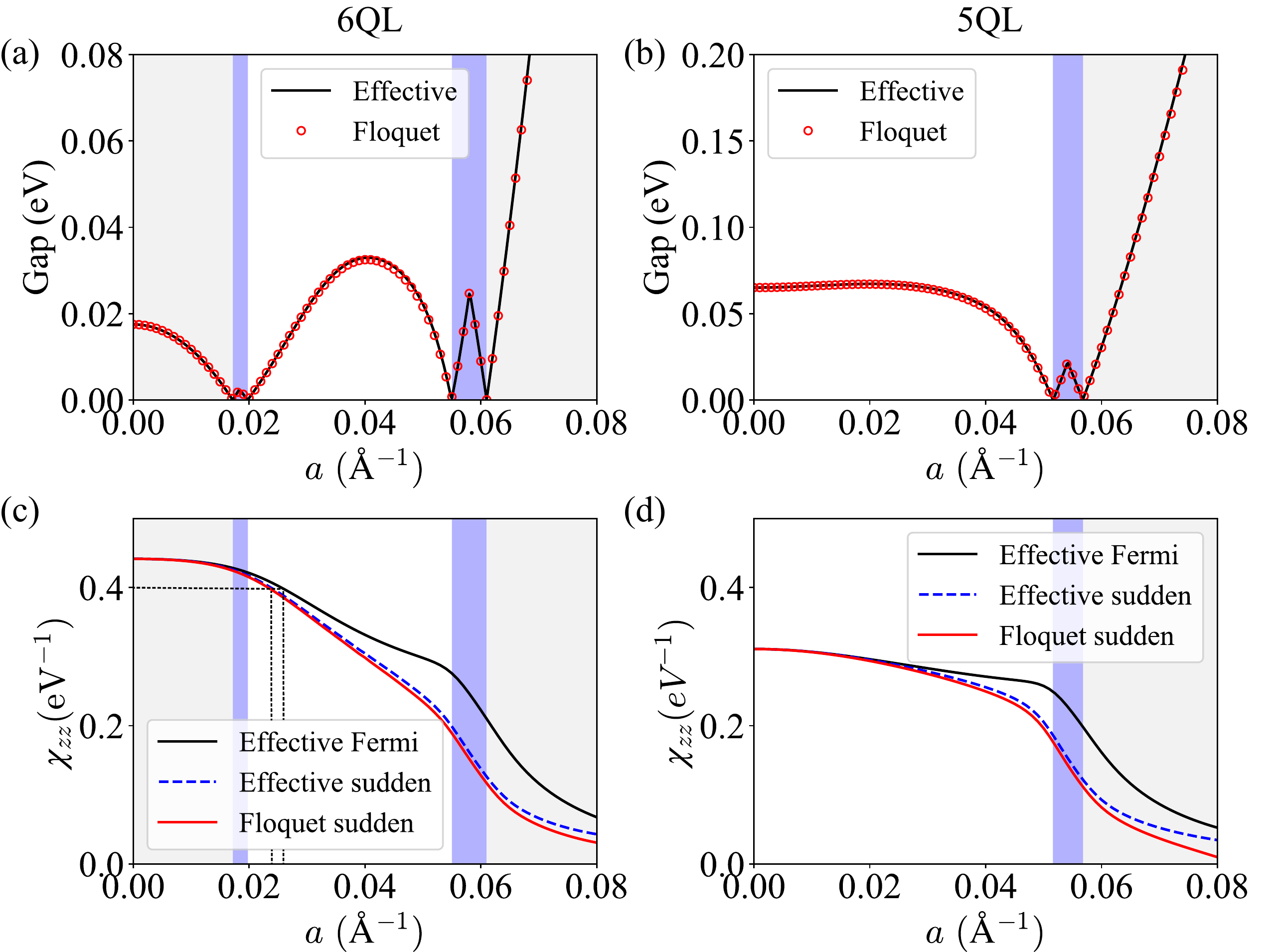}
\caption{(a,b) Floquet gap size changes from $(\mathrm{Sb}_{0.9}\mathrm{Bi}_{0.1})_2 \mathrm{Te}_3$ thin film model with 6QL (a) and 5QL (b). Black lines correspond to the high frequency expansion. Red lines are obtained by diagonalizing the infinite Floquet Hamiltonian with a cut-off on the frequency domain. (c,d) The change of magnetic susceptibilities with $a$ in MDTI thin films with 6QL (c) and 5QL (d). Black solid lines and blue dashed lines are for the high frequency expansion with different population approximations. Red curves are from diagonalization with sudden approximation. Background colors indicate topological phases and follow the convention in Fig. \ref{fig:2dphase}.} \label{fig:3d}
\end{figure}

To understand the effect induced by CPL in thin film model, we do similar calculations as those by using the effective Hamiltonian. Figure \ref{fig:3d} shows direct Floquet gap sizes and magnetic susceptibilities for $(\mathrm{Sb}_{0.9}\mathrm{Bi}_{0.1})_2 \mathrm{Te}_3$ thin films under CPL. Similar to results from simple model, light induced topological and magnetic PTs exist in intrinsic and doped thin films. A more complicated topological phase diagram is observed for 6QL when $a$ increases. They go through phases from NI$A$ to QAH, to NI$B$, to QAH again, and finally enter a NI$A$ phase. Such oscillatory crossover behaviour is dominated by QWSs \cite{SuppMater} and cannot be captured by the simple 2D effective model. For 5QL thin film, the topological phase transition driven by the light is similar to that shown in Fig. \ref{fig:2d_result}(a), suggesting its topological characters can be well described by the 2D effective Hamiltonian.

When we dope magnetic ions and keep its density to the value used in Fig. \ref{fig:2d_result}, the MDTI thin film with 6QL becomes FM while 5QL does not, which indicates that the 2D effective model is not well suited to describe the realistic magnetic properties for 5QL thin film. Their magnetic susceptibilities in the Floquet states decrease with the increase of light intensity and its slopes change differently in different topological phases. $a_c^{FM}$ in magnetic PTs also depends on the quantum quenching and does not correspond directly to the critical value of the topological PT. These results obtained from realistic thin film models are consistent with conclusions from the 2D effective model, confirming the validity of our physical picture.

%The former is with Floquet Fermi-Dirac distribution, and the latter is with SA.
%Magnetic critical point is marked by black dashed line.
%Three quantum well states are used.

\emph{Conclusion and Outlook}.-- Using the Floquet theory, we study topological and magnetic properties for TI and MDTI thin films under CPL. In the high frequency limit, topological PTs can be induced in TI thin films by increasing the intensity of the driving laser. Applying the same setup to MDTI thin films, we find magnetic PTs from FM to PM phases and their critical behaviors depend on the quantum quenchings that influence the non-equilibrium charge population. Based on previous discussions using the Floquet-Green's function method \cite{ono_nonequilibrium_2018}, we argue that the high frequency approximation is reasonable in a considerable large frequency regime, even when $\hbar\omega$ is comparable to $W$. On the other hand, when $\hbar\omega$ is smaller than $W$ with the resonant photon absorption, $\chi_{zz}^{Floq}$ still behaves similarly to the non-resonant case except in photon absorption regions \cite{SuppMater}. Thus, we are convinced that our conclusion could apply in a large parameter regime.

The phenomena proposed in this work are possible to be observed by different experimental setups. In pump-probe detection of the polar Kerr-rotation angle \cite{Kirilyuk2010}, the population effect could be described by the sudden approximation and the change of magnetism could be measured. Meanwhile, the Floquet Fermi-Dirac distribution is a good approximation for the transport measurement, such as the ultrafast anomalous Hall effect by laser-triggered photoconductive switches \cite{mciver2018light,Sato2019}. We expect that light driven topological and magnetic PTs can be realized in TI and MDTI thin films.

X. L. and W. D. acknowledge financial supports from the Ministry of Science and Technology of China (Grant No. 2016YFA0301001), the National Natural Science Foundation of China (Grant No. 11674188), and the Beijing Advanced Innovation Center for Future Chip (ICFC). A. R., H. H., and U. D. G. acknowledge financial supports from the European Research Council (ERC-2015-AdG-694097). P. T. acknowledges the received funding from the European Union Horizon 2020 research and innovation programme under the Marie Sklodowska-Curie grant agreement No 793609. The Flatiron Institute is a division of the Simons Foundation.

%%%%\bibliography{Cites}


\end{document}